\documentclass{moriond}

\usepackage{amsmath}
\usepackage{amsthm}

\usepackage{units}
\usepackage{mathrsfs}
\usepackage{amssymb}
\usepackage{slashed}
\usepackage{multirow}
\usepackage{mathtools}
\usepackage{nicefrac}
\usepackage{dsfont}
\usepackage{mathrsfs}
\usepackage{verbatim}
\usepackage{commath}
\usepackage{slashed}
\usepackage{ulem}
\usepackage{graphicx}
\usepackage{caption}

\newcommand{\cY}{\mathcal{Y}}%

\newcommand{\LL}{\mathscr{L}}%
\newcommand{\YE}{\mathcal{Y}_E}%
\newcommand{\YN}{\mathcal{Y}_N}%
\newcommand{\beq}{\begin{equation}}%
\newcommand{\eeq}{\end{equation}}%
\newcommand{\bea}{\begin{align}}%
\newcommand{\ena}{\end{align}}%

\newcommand{\hc}{{\rm h.c.}}

\begin{document}
\vspace*{4cm}
\title{GAUGING LEPTON FLAVOUR}

\author{ PABLO QU\'ILEZ LASANTA }

\address{ Departamento de F\'isica Te\'orica UAM and Instituto de F\'isica Te\'orica UAM-CSIC, \\ Universidad Aut\'onoma de Madrid, Cantoblanco, 28049, Spain}

\maketitle\abstracts{
The flavour lepton symmetry is gauged for both the Standard Model flavor group and its extension with three degenerate right-handed neutrinos. As a consequence of anomaly cancellation, exotic leptons are introduced which induce a see-saw mechanism that generates the masses of the SM leptons. The model is compared with  Minimal Flavour Violation, the difference between the low-energy effective operators resulting is identified, as well as the distinctive experimental signals, particularly promising in the $\mu-\tau$ sector.}
\section{Motivation}

The origin of the flavour structure of the elementary particles composing the visible universe remains as one of the most fundamental questions in Particle Physics. Flavour is also central constraining beyond the Standard Model theories (BSM) attempting to solve the electroweak hierarchy problem since they typically imply unacceptable consequences in the flavour sector: this is known as the flavour problem. In order to solve this problem, it is suggesting to consider theories that explain the flavour puzzle in terms of a symmetry principle.

Inspired by this idea, the promotion of the lepton flavour symmetry to a gauge symmetry is considered for the SM flavour group, $SU(3)_{\ell}\times SU(3)_E$, and its extension with 3 degenerate right-handed neutrinos~\cite{Alonso:2016onw}, $SU(3)_{\ell}\times SU(3)_E\times SO(3)_N$, extending previous work on the quark sector~\cite{Grinstein:2010ve}.

\section{Gauged Lepton Flavour Standard Model: $SU(3)_{\ell}\times SU(3)_E$}
The leptonic global flavour symmetry to be gauged is that exhibited by the SM in the absence of Yukawa couplings~\cite{Chivukula:1987py}: $SU(3)_{\ell}\times SU(3)_E$. Anomaly cancellation of this non-abelian symmetry is accomplished by the addition to the Lagrangian of three extra fermion species,  denoted  by $\mathcal{E}_R$, $\mathcal{E}_L$, and $\mathcal{N}_R$ whose quantum numbers are shown in Table~\ref{GLFSMPsi}. In addition, two scalar \textit{flavon} fields, $\YE$ and $\YN$, are introduced to assure flavour invariance of the Yukawa couplings. As we will see their vevs will be related with the inverse of the mass matrices of the leptons.

\begin{table}[h!]
\centering
\begin{tabular}{c|cccc} 
&  $SU(2)_L $ & $U(1)_Y$ & $SU(3)_\ell$ & $SU(3)_E$\\ 
\hline 
\hline 
$\ell_L \equiv(\nu_L\,, e_L)$  	& 2 			& $-1/2$		& 3			& 1\\
$e_R$   					& 1  			& $-1$		& 1			& 3 \\
\hline
${\mathcal{E}_R}$   			&   1  		& $-1$ 		& 3			&1\\
$\mathcal{E}_L$    			&  1  			& $-1$ 		& 1			&3\\
$\mathcal{N}_R$   					&  1 			& 0 			& 3 			&1\\
\hline
$\YE$   					&   1  		& 0 			& $\bar3$		& $3$\\
$\YN$    					&  1  			& 0  			& $\bar6$ 		& $1$\\
\end{tabular}
\caption{\it Transformation properties of SM fields, of mirror fields and of flavons under the SM and $SU(3)_\ell\times SU(3)_E$.}
\label{GLFSMPsi}
\end{table}
The most general renormalizable lagrangian with those fields and symmetries  for the Yukawa and mass terms reads:
\beq
\begin{aligned}
\mathscr{L}_{Y}=&\,\lambda_E\, \overline \ell_L\, H \,\mathcal{E}_R+\mu_E\,\overline{\mathcal{E}}_L\, e_R  +\lambda_{\mathcal E}\,\overline{\mathcal{E}}_L \, \mathcal{Y}_E \,\mathcal{E}_R+\hc\\
&+\lambda_\nu\,\overline \ell_L\, \tilde H \,\mathcal{N}_R +\frac{\lambda_N}{2}\,\overline {{\mathcal{N}_R}^c}\, \mathcal{Y}_N\, \mathcal{N}_R+\hc\,,
\end{aligned}
\label{YLAGLFSM}
\eeq

When the flavon fields $\YE$ and $\YN$ develop a vacuum expectation value (vev), the flavour symmetry breaking is trigered and fermion masses are generated. The dynamics of flavour breaking is encoded in the scalar potential, whose minima will determine the vevs of the flavons that have been studied in Refs.~\cite{Alonso:2012fy,Alonso:2013nca,Alonso:2013mca}. Although a dynamical justification for all fermion masses and mixings is still lacking it has been shown that the potential minima lead to no mixing in the quark sector and large mixings and majorana phases in the lepton sector if the degenerate right-handed neutrinos are introduced (see Section~\ref{GaugedSeesaw}).

The mass matrices generated by the Lagrangian in Eq.~\ref{YLAGLFSM} suggest a Seesaw-like pattern for both charged and neutral leptons. In the limit were $\mathcal{Y}_E\gg \mu_E\,,v$, the masses of the exotic fermions $\mathcal{M}_{\mathcal{E}}$, $\mathcal{M}_{N}$ and SM leptons $m_\ell$, $m_\nu$ read:
\begin{align}
 \mathcal{M}_{\mathcal{E}}&=\lambda_{\mathcal{E}} \YE,
& \mathcal{M}_{N}&= \lambda_N\YN,\\
m_\ell\,\,&= \frac{v}{\sqrt{2}}\frac{\lambda_E\, \mu_E}{\lambda_{\mathcal{E}}\YE} ,
& m_\nu\,\,&= \frac{v^2}{2}  \frac{\lambda^2_\nu}{\lambda_{N}\YN} 
\label{SMLMass}
\end{align}
Due to the seesaw-like mechanism that arises for all leptons, the SM fermion masses are inversely proportional to the masses of the exotic fermions. From the phenomenological perspective this inverse proportionality  is really interesting since the deviations from the SM predictions are more important for $\tau$ - related observables, which are less constrained. 

In general, the expected phenomenological  signals  of this model are flavour-conserving, and include charged-lepton universality violation and non-unitarity of the PMNS matrix. Furthermore, the first particles awaiting discovery would be a tau mirror lepton and $SU(3)_E$ gauge bosons which mediate $\mu_R-\tau_R$ transitions.


\section{Gauged Lepton Flavour Seesaw Model: $SU(3)_{\ell}\times SU(3)_E\times SO(3)_N$} \label{GaugedSeesaw}

Secondly we have considered the gauging of the flavour symmetry of the type I Seesaw theory with three degenerate right-handed neutrinos $N_R$. In this context, the maximal non-abelian flavour symmetry group of the Lagrangian in the limit of vanishing masses is $SU(3)_\ell\times SU(3)_E\times SO(3)_N$. The fermionic field content that needs to be added in order to cancel gauge anomalies is identical to that in the previous model \footnote{triangle diagrams cancel for $SO(3)_N$ and the $N_R$ fermions are singlets under the SM gauge symmetry.} while both scalar flavon fields transform in the bifundamental of the flavour group \footnote{unlike the previous case where $\mathcal{Y}_N$ transforms in the conjugate symmetric representation of $SU(3)_\ell$} (see Table~\ref{SPTLFSeM}).
\begin{table}[h]
\centering
\begin{tabular}{c|ccccc} 
&  $SU(2)_L $ & $U(1)_Y$ & $SU(3)_\ell$ & $SU(3)_E$&$SO(3)_N$\\ 
\hline
\hline \
$N_R$  &   1  & 0 & 1 & 1 & 3\\
${\mathcal{Y}_N}$    &  1  & 0  & $\bar 3$ & 1 & 3\\
\end{tabular}
\caption{\it Transformation properties of the fields that differ from the first model under the full gauge group.}
\label{SPTLFSeM}
\end{table}

 The Yukawa interactions and Majorana mass terms of this model read:
\beq
\begin{split}
\LL_{Y}=&\lambda_E\, \overline \ell_L\, H \,{\mathcal{E}_R} +\mu_E\,\overline{\mathcal{E}}_L \,e_R +\lambda_{\mathcal E}\,\overline{\mathcal{E}}_L \, \mathcal{Y}_E \,{\mathcal{E}_R} \\
&+\lambda_\nu\,\overline \ell_L\, \tilde H \,\mathcal{N}_R +\lambda_N\,\overline {N_R^c}\, \mathcal{Y}_N \,\mathcal{N}_R+\dfrac{\mu_{LN}}{2}\, \overline {{N}_R}^c  N_R+\mathrm{h.c.},
\label{LagISM}
\end{split}
\eeq

The particle spectrum and phenomenology of the charged lepton sector matches the description given in the previous section, while for the $SU(3)_\ell$ and $ SO(3)_N$  sectors the spectrum and phenomenology will now depend on three fundamental scales: the vevs of $\YE$ and $\mathcal{Y}_N$ and the lepton number parameter $\mu_{LN}$. From now on we will focus on the limit $\cY_N\gg \mu_{LN}$ since in the opposite $\cY_N\ll \mu_{LN}$ we recover the phenomenology of the previous section.

  In particular, the neutral fermion mass matrix is the typical of $\bold{inverse\,\, Seesaw}$ scenarios:
\beq
\frac{1}{2}\left(
\begin{array}{ccc}
0&\lambda_\nu v/\sqrt{2}&0\\
\lambda_\nu v/\sqrt{2}&0 &\lambda_N \mathcal{Y}_N^T\\
0&\lambda_N \mathcal{Y}_N &\,\,\mu_{LN}
\end{array}
\right)\, \, + \textrm{h.c.},
\label{massmatrices2}
\eeq
The masses of the exotic neutral leptons $\mathcal{M}_N$ and neutrinos $m_{\nu}$ read:
\begin{align}
\mathcal{M}_N&\simeq\lambda_N\cY_N\,, &m_\nu \simeq &\dfrac{v^2}{2} \dfrac{{\lambda_\nu}^2}{\lambda_N^2}\, \dfrac{1}{\cY_N} \mu_{LN} \dfrac{1}{\cY_N^T}\,.
\label{MassND}
\end{align}

The main phenomenological difference of this model with respect to the previous one is that now we can expect not only LUV signals but also Lepton Flavour Violation (LFV), precisely because the LN parameter $\mu_{LN}$ and lepton flavour violation scale $\norm{\cY_N}$ are independent and the latter is not strongly constrained by the tiny value of light neutrino masses.


\section{Comparison with Minimal Lepton Flavour Violation}

In this work we have gauged  the flavour symmetry of the SM  and of the type I Seesaw Lagrangian inspired by the idea of solving the flavour problem through a symmetry principle, therefore it is interesting to study whether the resulting low-energy phenomenology is compatible with that expected by other attempts like  Minimal Lepton Flavour Violation (MLFV)~\cite{Cirigliano:2005ck}. 

Minimal flavour violation (MFV) is an effective aproach that describes the low energy  effects of a class models that are not afflicted by the flavour problem by imposing to the BSM to respect the flavour symmetry, plus the simple assumption that at low-energies Yukawa couplings are the only source of flavour.

\begin{minipage}{0.32\linewidth}
\hspace{10pt}\centerline{\includegraphics[width=1.25\textwidth]{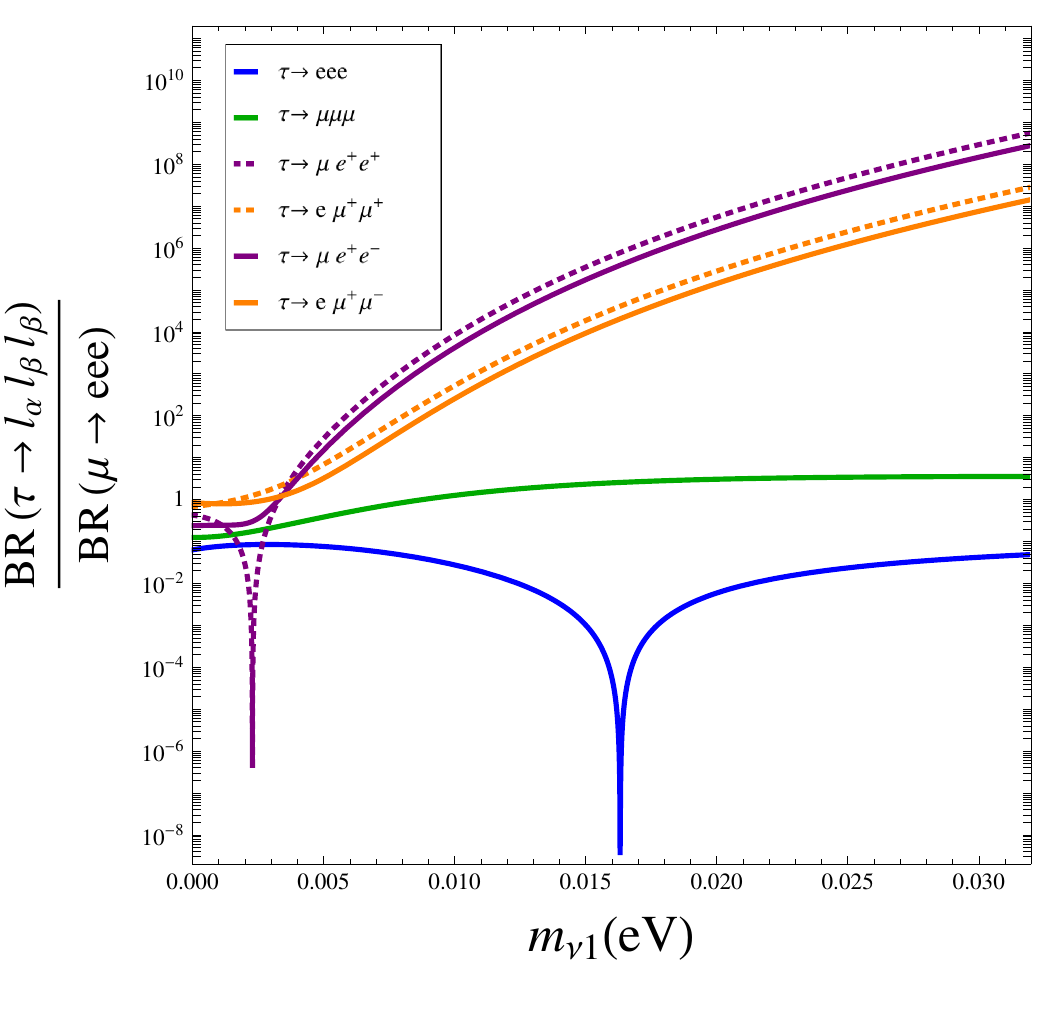}}
\end{minipage}
\hspace{2.5cm}
\begin{minipage}{0.33\linewidth}
\hspace{10pt}\centerline{\includegraphics[width=1.25\textwidth]{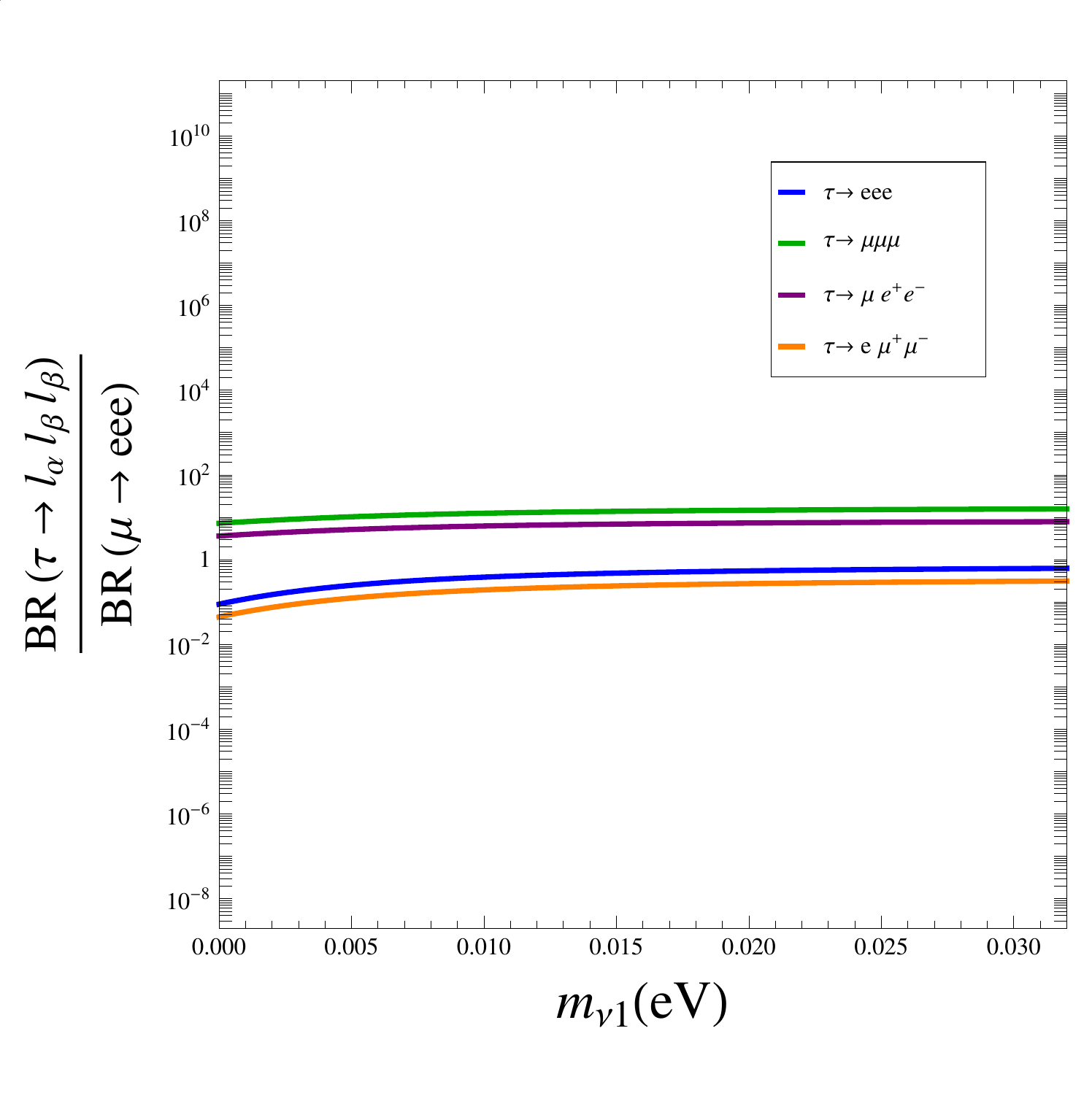}}
\end{minipage}\\
\,\, {\small Figure 1 -} {\footnotesize Comparison between the gauged-flavour type-I Seesaw scenario for $\YN\gg\YE$ and MLFV in a CP-even case: branching ratios for the different lepton rare decays over that for $\mu\to eee$, for neutrino normal ordering.}

The low-energy effective Lagrangian of our gauged-flavour models is, by construction, formally invariant under the spurion analysis of  MLFV; however the analytic dependence on the scalar fields does not always match that in  the original formulation of MLFV and subsequent works~\cite{Cirigliano:2006su,Davidson:2006bd,Gavela:2009cd,Alonso:2011jd}.
In particular we found that, while the effective operators that arise due to exotic fermion exchange resemble those of MLFV, the operators resulting from the flavour gauge boson exchange have a more complex structure and do not correspond to any operator in the classical formulation of MLFV, therefore leading to distinctive phenomenological signals.

 As an illustration,  the  predicted branching ratios for various $l_\alpha\to l_\beta l^+_\rho l^-_\kappa$ processes are compared  in Fig. 1. The main difference within these Lepton Flavour Violating decays is that the gauged flavour model predicts processes that violate lepton flavour by two units at leading order (e.g., $\tau\rightarrow \mu e^+ e^+$ 
and $\tau \rightarrow e \mu^+ \mu^+$) while they are absent in the MLFV case since they are suppressed by higher-order spurion insertions.

\section{Conclusions}
We have considered the gauging of leptonic global flavour symmetries that the SM Lagrangian or its fermionic Seesaw extension exhibit in the limit of massless SM leptons. It is remarkable that the gauge anomaly cancellation conditions 
point to a  universal Seesaw pattern for both charged and neutral leptons. In the neutrino sector the gauging of the SM flavour leads to the type I Seesaw  while the gauging of the Seesaw flavour symmetry generates an inverse Seesaw scenario. 

Lepton Universality Violation signals are expected for both the gauged SM-flavour and the gauged Seesaw flavour models  and, in the second case, flavour violating transitions among charged leptons could arise.  As a consequence of the Seesaw pattern the main expected signals tend to involve the heavier SM leptons whose interactions are less constrained by present data.

The phenomenology has been compared with that of Minimal Lepton Flavour Violation and it was shown that flavour gauge bosons may induce distinctive charged lepton transitions with respect to those expected in MLFV while the mirror lepton effects mimic those of MLFV.

\section*{Acknowledgements}
\vspace{-6pt}
My work is funded by Fundacion La Caixa under "La Caixa-Severo Ochoa" predoctoral grant. I acknowledge partial financial support by the European Union through the INVISIBLESPLUS, by the Horizon2020-MSCA-ITN-2015//674896-ELUSIVES, by CiCYT through the project FPA2016-78645-P, and by the Spanish MINECO through the Centro de excelencia Severo Ochoa Program under grant SEV-2012-0249.  A special thanks to all the collaborators from Madrid (M.B. Gavela, E. Fernandez-Martinez and L.Merlo) and San Diego (R. Alonso and B. Grinstein). I thank the the organisers for the kind invitation and for such an interesting conference.

\section*{References}


\bibliography{biblio}{}
\bibliographystyle{unsrt}
\end{document}